\documentstyle[12pt]{article}

\def\bar{\overline}
\def\hat{\widehat}
\def\*{\star}
\def\({\left(}		\def\BL{\Bigr(}
\def\){\right)}		\def\BR{\Bigr)}
\def\[{\left[}		\def\BBL{\Bigr[}
\def\]{\right]}		\def\BBR{\Bigr]}

\def\frac#1#2{{#1 \over #2}}		
\def\inv#1{{1 \over #1}}
\def\half{{1 \over 2}}
\def\d{\partial}

\def\ket#1{ | #1 \rangle}
\def\bra#1{ \langle #1 |}

\def\2pi{\hbox{$2\pi i$}}

\def\dsl{\raise.15ex\hbox{/}\kern-.57em\partial}
\def\Dsl{\,\raise.15ex\hbox{/}\mkern-.13.5mu D}
\def\th{\theta}

\def\ep{\epsilon}
	\def\La{\Lambda}
\def\de{\delta}		\def\De{\Delta}
\def\om{\omega}		\def\Om{\Omega}
\def\sig{\sigma}	

	\def\CE{{\cal E}}	\def\CF{{\cal F}}

\def\CM{{\cal M}}		
		
\def\CS{{\cal S}}

\font\numbers=cmss12
\font\upright=cmu10 scaled\magstep1
\def\stroke{\vrule height8pt width0.4pt depth-0.1pt}
\def\topfleck{\vrule height8pt width0.5pt depth-5.9pt}
\def\botfleck{\vrule height2pt width0.5pt depth0.1pt}
\def\Zmath{\vcenter{\hbox{\numbers\rlap{\rlap{Z}\kern 0.8pt\topfleck}\kern
2.2pt
                   \rlap Z\kern 6pt\botfleck\kern 1pt}}}
\def\Qmath{\vcenter{\hbox{\upright\rlap{\rlap{Q}\kern
                   3.8pt\stroke}\phantom{Q}}}}
\def\Nmath{\vcenter{\hbox{\upright\rlap{I}\kern 1.7pt N}}}
\def\Cmath{\vcenter{\hbox{\upright\rlap{\rlap{C}\kern
                   3.8pt\stroke}\phantom{C}}}}
\def\Rmath{\vcenter{\hbox{\upright\rlap{I}\kern 1.7pt R}}}
\def\Z{\ifmmode\Zmath\else$\Zmath$\fi}
\def\Q{\ifmmode\Qmath\else$\Qmath$\fi}
\def\N{\ifmmode\Nmath\else$\Nmath$\fi}
\def\C{\ifmmode\Cmath\else$\Cmath$\fi}
\def\R{\ifmmode\Rmath\else$\Rmath$\fi}
\def\eqn#1{\begin{eqnarray} #1 \end{eqnarray} }
\def\non{\nonumber}

\begin{document}
\hsize37truepc\vsize61truepc
\hoffset=-.5truein\voffset=-0.8truein
\setlength{\baselineskip}{17pt plus 1pt minus 1pt}
\setlength{\textheight}{25.5cm}
\vphantom{0}\vskip1.1truein
\leftline{{\Large{A One Dimensional Ideal Gas of Spinons,}}}
\leftline{ or}
\leftline{Some Exact Results on the XXX Spin Chain
with Long Range Interaction.}
\vskip.65truein

\hbox{\obeylines\baselineskip12pt\parskip0pt\parindent0pt\hskip1.1truein
\vbox{D. Bernard, V. Pasquier and D. Serban,
\vskip.1truein
Service de Physique Th\'eorique de Saclay \footnote{Laboratoire de la direction
des sciences de la mati\`ere du commissariat \`a l'\'energie atomique.},
F-91191, Gif-sur-Yvette, France.
}}
\vskip .65truein
\noindent
%
Abstract: We describe a few properties of the XXX spin chain
with long range interaction. The plan of these notes is~:\\
1 --- The Hamiltonian.\\
2 --- Symmetry of the model.\\
3 --- The irreducible multiplets.\\
4 --- The spectrum.\\
5 --- Wave functions and statistics.\\
6 --- The spinon description.\\
7 --- The thermodynamics.\\

\vskip.65truein

\noindent {\bf Introduction.}
The XXX spin chain with long range interaction is a variant
of the spin half Heisenberg chain, with exchange inversely
proportional to the square distance between the spins.
It possesses the remarkable properties that its spectrum is
additive and that the elementary excitations are spin
half objects obeying a half-fractional statistics
intermediate between bosons and fermions.
In this sense, it gives a model for an ideal gas
of particles with fractional statistics.
The model is gapless; its low energy properties belong
to the same universality class as the Heisenberg model,
and are described by the level one $su(2)$ WZW conformal
field theory.

\bigskip
{\bf Acknowledgements:} It is a pleasure to thank Olivier
Babelon, Michel Gaudin and Duncan Haldane for stimulating discussions.

\section { The Hamiltonian.}
The Hamiltonian of the trigonometric isotropic spin chain
with long range interaction is given by \cite{Ha88,Sha88}~:
\eqn{ H\ =\ \({\frac{\pi}{N}}\)^2\sum_{i\not= j}\
\frac{(P_{ij}-1) }{\({ 2\sin\frac{\pi(i-j)}{2N} }\)^2 }
\label{EAa} }
where $P_{ij}$ is the operator which exchange the spins
at the sites $i$ and $j$.
We restrict ourselves to the $su(2)$ case, in which case
the spin variables can only take two values: $\sigma_i= \pm$.
The sum is over all the distinct pairs of sites labeled by
integers $i,j,\cdots$ ranging from $1$ to $N$.

The spectrum of (\ref{EAa}), which has been conjectured by Haldane
\cite{Ha91},
 possesses a remarkable additivity property
as well as a rich degeneracy. It can be described as follow.
To each eigenstate multiplet is associated a set of
rapidities $\{m_p\}$ which are non-consecutive integers
ranging from $1$ to $(N-1)$.
The energy of an eigenstate $\ket{\{m_p\}}$ with rapidities $\{m_p\}$ is:
\eqn{
H \ket{\{m_p\}}\ =\ \({\sum_p \ep(m_p)}\) \ket{\{m_p\}}
\qquad {\rm with}\quad \ep(m)= \({\frac{\pi}{N}}\)^2 m(m-N)
\label{EAb} }
The degeneracy of the multiplet with rapidities $\{m_p\}$
 is described by its $su(2)$ representation content as follows.
Encode the rapidities in a sequence of $(N-1)$ labels $0$ or $1$
in which the $1$'s indicate the position of the rapidities;
add two $0$'s at both extremities of the sequence which now has
length $(N+1)$. Since the rapidities are never equal nor differ by
a unit, two labels $1$ cannot be adjacent.
A sequence can be decomposed into the product of elementary
motifs. A motif is a series of $Q$ consecutives $0$'s, and it corresponds
to a spin $\frac{Q-1}{2}$ representation of $su(2)$. The representation
content of a sequence is then the tensor product of its motifs.

The degeneracy of the spectrum can also be described in terms of path,
in a way surprisingly similar to the path description of the
six-vertex corner transfer matrix \cite{DJKMO}.

\section {Symmetry of the model.}
The symmetry algebra responsible for the degeneracy of the
model was identified as the $su(2)$ Yangian \cite{nous}.
A Yangian is an infinite dimensional associative
algebra generated by elements $T_n^{ab}$, with $n$
a positive integer, and $a,b=\pm$ in the $su(2)$ case.
These generators satisfy quadratic relations which can be arranged
into an Yang-Baxter equation by introducing the transfer matrix
$T(x)$, with matrix elements,
$T^{ab}(x) = \de^{ab} + \sum_{n\geq 0}x^{-n-1} T^{ab}_n$.
The commutation relations then take the following form
\cite{Dr87,Skly}~:
\eqn{
R(x-y)\ (1\otimes T(x))\ (T(y)\otimes1)\ =\
(T(y)\otimes 1)\ (1\otimes T(x))\ R(x-y)
\label{EBa} }
The matrix $R(x)$ is the solution of the Yang-Baxter equation given by:
$R(x) = x + P$, where $P$ is the permutation operator
which exchanges the two auxiliary spaces.
The transfer matrix was constructed in \cite{nousbis} . Its expression is~:
\eqn{
 T^{ab}(x)= \de^{ab} +\sum_{i,j=1}^N X_i^{ab}\Bigl({\inv{x-L} }\Bigr)_{ij}
\label{EBb} }
with $L_{ij}=(1-\de_{ij})\th_{ij}P_{ij}$,
$\theta_{ij}=z_i/z_{ij}$ with $z_{ij}=z_i-z_j$, and $X_i^{ab}$ is the canonical
matrix $\ket{a}\bra{b}$ acting on the $i^{th}$ spin only.
The transfer matrix (\ref{EBb}) form a representation of the exchange
algebra (\ref{EBa}) for any values of the complex numbers $z_j$.
The center of the $su(2)$ Yangian algebra
(\ref{EBa}) is generated by the so-called
quantum determinant $Det_qT(x)$ defined by \cite{Kor}:
\eqn{
Det_qT(x) = T_{--}(x-1)T_{++}(x) - T_{-+}(x-1)T_{+-}(x)
\label{EBc} }
In the representation (\ref{EBb}) , the quantum determinant is a pure number
 for any values of the $z_j$'s given by~:
\eqn{
 Det_q\ T(x)\ =\ 1 + \sum_{i,j=1}^N \Bigl({ \inv{x-\Theta}}\Bigr)_{ij}
 =\ \frac{ \De_N(x+1) }{\De_N(x) }
\label{EBd} }
with $\De_N(x) $ the characteristic polynomial of the
$N\times N$ matrix $\Theta$ with entries $\th_{ij}$:
$\De_{N}(x)= {\rm det}(x-\Theta)$.

The trigonometric spin chain corresponds to $z_j=\om^j$ with
$\om$ a primitive $N^{th}$ root of the unity. For these values
of $z_j$, the tranfer matrix (\ref{EBb}) commutes with
the Hamiltonian (\ref{EAa}).
For $z_j=\om^j$, the matrix $\Theta$ can be diagonalized
giving the following expression for $\De_N$~:
\eqn{
\De_N(x)= \prod_{j=1}^N \bigl( x+\frac{N+1}{2} -j \bigr)
\label{EBz} }

\section {The irreducible multiplets.}
Solving the model consists in finding all the irreducible
components of the Yangian symmetry algebra and computing
the energy in each of these blocks. For the values of the $z_j$'s
 induced by the spin chain, $z_j=\om^j$, the representation (\ref{EBb})
is reducible. It is completely
reducible since the transfer matrix is hermitic:
${t^{ab}_n}\,^{\dag}=t^{ba}_n$.
Each irreducible sub-representation
possesses a unique highest weight (h.w.) vector $\ket{\La}$ which
is annihilated by $T_{+-}(x)$ and which is an eigenvector of the
diagonal components $T_{\pm\pm}(x)$ of the transfer matrix~:
\eqn{
 T(x)\ket{\La} = \pmatrix{ t_{++}(x) & 0 \cr
		           \star & t_{--}(x) } \ket{\La}
\non }
Here, $t_{\pm\pm}(x)$ are rational functions in $x$, but not operators.
Since the quantum determinant (\ref{EBb}) take the same value in any of the
irreducible block, these two functions are related by~:
\eqn{
\frac{\De_N(x+1)}{\De_N(x)} = t_{--}(x-1) t_{++}(x)
\non }
Hence, only one of them, say $t_{--}(x)$, is independent.
It uniquely characterizes the $su(2)$ Yangian representation.
We therefore have to compute all the functions $t_{--}(x)$
arising from the decomposition of the Yangian representation
induced by the spin chain, but also to identify
the h.w. vectors in order to be able to compute
the energy spectrum.

Obviously, the ferromagnetic vacuum $\ket{\Om}=\ket{++\cdots++}$
is a h.w. vector: the corresponding  $t_{--}(x)$ is one,
and the energy is zero.
The h.w. vectors in the one-magnon sector
are $\ket{m}=\sum_j\om^{mj}\sig_j^-\ket{\Om}$, with, $1\leq m\leq(N-1)$:
the corresponding eigenvalue is $t_{--}(x)=\frac{P_1(x+1)}{P_1(x)}$,
with $P_1(x)=(x+\frac{N+1}{2}-m)$,
and the one-magnon energy is $\ep(m)=\({\frac{\pi}{N}}\)^2m(m-N)$.

In order to determine all the highest weight vectors, we
decompose the hilbert space into subspaces of fixed magnon
number. A $M$-magnon state $\ket{\Psi}$ has $M$ spin reversed~:
\eqn{ \ket{\Psi}\ = \sum_{n_1,\cdots,n_M}\ \psi_{n_1,\cdots,n_M}\
\sig^-_{n_1}\cdots \sig^-_{n_M} \ket{\Om}
\label{ECa} }
where $\sig^a_n$ denote the Pauli matrices acting on the spin
located on the site $n$. By construction, the coefficients
$\psi_{n_1,\cdots,n_M}$ of the $M$-magnon wave functions are symmetric
in their indices.  The wave function coefficients are unspecified
for two coincident indices $\psi_{\cdots,n,\cdots,n, \cdots}$.
By convention, we choose these coefficients to be zero.

Since these indices range from $1$ to $N$,
to any $M$-magnon state is associated a symmetric polynomial
$\Psi(z_1,\cdots,z_M)$\ in $M$ variables of degree less than $(N-1)$
such that $\Psi(\om^{n_1},\cdots,\om^{n_M})=\psi_{n_1,\cdots,n_M}$.
In the following, we restrict ourselves to the class of magnon states
deriving from polynomials of the following form~:
\eqn{
\Psi(z_1,\cdots,z_M)= \prod_{p<q}(z_p-z_q)^2
	R(z_1,\cdots,z_M)
\label{EZaa} }
with $ R(z_1,\cdots,z_M)$ a symmetric polynomial of degree
less than $(N-2M+1)$. This class of states does not include
all the states of the spin chain but, as we will see,
all the highest weight vectors are in this class.

As explained in the Appendix, the operators $T_{--}(x)$
and $T_{+-}(x)$ act on this class of states. Therefore,
the action of these operators on these magnon states
induces an action on the polynomials. As shown in the
Appendix, we find~:
\eqn{
T_{--}(x)\  \Psi(z_\cdot)\
= \({1 + \sum_{p=1}^M\inv{x+\frac{N+1}{2}-D_p} }\) \Psi(z_.)
\label{EDh}}
and
\eqn{
T_{+-}(x)\  \Psi(z_\cdot)
= \({1 + \sum_{p=1}^M\inv{x+\frac{N+1}{2}-D_p} }\) \Psi(z_1=0,z_.)
\label{EDhh}}

Here we have introduced differential operators
$D_p$ which have recently been proved
useful in the Calogero-Sutherland model
\cite{Heck,Poly,Cherd,nousbis}.
For the following, we also need another set of differential
operators $\hat D_p$.  Both are defined by~:
\eqn{
D_p &=& z_p\d_{z_p} + \sum_{p\not= q} \theta_{pq} K_{pq} \non\\
\hat D_p &=& z_p\d_{z_p} + \sum_{q>p} \theta_{pq} K_{pq}
- \sum_{q<p} \theta_{qp} K_{pq}
\label{ECd} }
Here, $\theta_{pq}=z_p/z_{pq}$ and,
 the operator $K_{pq}$ exchanges the positions: $K_{pq}z_p=z_qK_{pq}$.
The differential $D_p$ are covariant under permutation of
the positions, $K_{pq} D_p = D_q K_{pq}$, whereas the $\hat D_p $'s
are not. On the other hand, the differentials $\hat D_p$ commute,
$\BBL\hat D_p ,\hat D_q\BBR = 0$, but the differentials $D_p$'s does not:
$\BBL D_p , D_q \BBR = (D_p-D_q) K_{pq}$.
The sum of the $m^{th}$ powers of both differentials form
two sets of commuting operators. However, they are not independent
thanks to the following relation:
\eqn{
 1 +\sum_{p=1}^M \inv{x-D_p} = \({ 1 + \inv{x-\hat D_1}}\)
 \cdots \({ 1 + \inv{x-\hat D_M}}\)
\label{ECf} }
In Eq. (\ref{ECf}) it is understood that the operators are acting
on functions symmetric in their arguments.

{}From eq. (\ref{EDhh}), we learn that the highest weight
vectors correspond to polynomials $\Psi(z)$ with $R(z)$
given by~:
\eqn{
R(z_1,\cdots,z_M)= (\prod_{p=1}^Mz_p)  \phi(z_1,\cdots,z_M)
\label{EDe} }
with $ \phi(z_1,\cdots,z_M)$ a symmetric polynomial of degree
less than $(N-2M)$.
Thus, in the $M$-magnon sector, there are
$\frac{(N-M)!}{M!(N-2M)!}$ independent highest weight vectors.

{}From eq. (\ref{EDh}), we learn that
the eigenfunctions of $T_{--}(x)$ are the eigenvectors
of the commuting hamiltonians of the Calogero-Sutherland model,
or equivalently of the commuting operators $\hat D_p$.
The symmetric eigenfunctions
$\Psi^{\{m_p\}}(z_\cdot)$ with,
\eqn{\sum_p (\hat D_p)^n\ \Psi^{\{m_p\}}(z_\cdot)
= \sum_p m_p^n\ \Psi^{\{m_p\}}(z_\cdot),
\non }
are polynomials with degree between $1$ and  $(N-1)$ if
$1\leq m_p \leq (N-1)$.
Hence, the M-magnon highest weight vectors $\ket{\{m_p\}}$,
with wave function given by $\Psi^{\{m_p\}}(z_\cdot)$,
are labeled by sets of M integers $\{m_1,\cdots,m_M\}$.
Due to the Vandermond prefactor in eq.(\ref{EZaa}),
these integers never coincide nor differ by a unit.
Using the factorisation relation (\ref{ECf}), one find
that the value of $t_{--}(x)$ for these highest weight
vectors are~:
\eqn{
t_{--}(x)=\frac{P_1(x+1)}{P_1(x)}
\quad{\rm with}\quad
P_1(x) = \prod_{p=1}^M(x+\frac{N+1}{2}-m_p)
\label{EDi} }

The dimension of the irreducible multiplets are encoded in the transfer
matrix eigenvalues. The eigenvalues $t_{--}(x)$ are given by
eq. (\ref{EDi}).
The remaining  eigenvalues $t_{++}(x)$ are computed from
the relation (\ref{EBb}). They can also be written in a product form.
The result is~:
\eqn{
T(x)\ket{\{m_p\}} = \frac{P_1(x+1)}{P_1(x)}
	\pmatrix{ \frac{P_0(x+1)}{P_0(x)} & 0\cr
				~ & ~ \cr
			\star & 1 \cr} \ket{\{m_p\}}
\label{EDk} }
with $P_1(x)$ given in eq.(\ref{EDi}), and
$P_0(x)$ and $P_1(x)$ factorize $\Delta_N(x)$~:
\eqn{\Delta_N(x) = P_0(x)\ P_1(x)P_1(x-1).
\label{EDj} }
One can check that the factorization equation (\ref{EDj}) admits
solutions only if the roots of $P_1(x)$ are not adjacent.
This provides one way to recover the rapidities  selection rules.

Let us decompose the sequence of rapidities $\{m_p\}$ in elementary motifs
as explained in Section 1. To each motif of length $Q$,
we associate a canonical transfer matrix defined by~:
\eqn{
 T^{ab}_{motif}(x) = \de^{ab} + \frac{S^{ab}}{x-x_0}
\label{EDn}}
where $S^{ab}$ are the matrices forming the spin $\frac{Q-1}{2}$
representation of $su(2)$ and $x_0$ is the position of the most
left label $0$ of the motif. It is easy to check that the matrix
(\ref{EDn}) satisfy the commutation relations (\ref{EBa}) .
The representation induced by the transfer matrix (\ref{EDk}) is then
seen to be equivalent to the irreducible tensor product of
the transfer matrices associated to each motifs:
\eqn{
T(x) \cong \bigotimes_{motifs}\ T_{motif}(x+\frac{N+1}{2})
\label{EDl} }
This is proved by comparing the eigenvalues of the diagonal
transfer matrix elements on the h.w. vector.
Therefore, we find that the multiplet of a rapidity sequence
is the tensor product of each of its motifs.

We have found one (and only one) irreducible representation
for each rapidity sequence. Their direct sum is a vector
space of dimension $2^N$. Therefore, it fills the Hilbert space
of the spin chain, and there is no other irreducible multiplet.

\section {The spectrum.}
Since all the irreducible multiplets are now identified,
finding the spectrum consists in computing the action of
the Hamiltonian on the highest weight vectors.
The hamiltonian (\ref{EAa}) is $su(2)$ invariant, hence it acts on
the $M$-magnon subspace. In the magnon basis,
this action is~:
\eqn{
(H\psi)_{n_1,\cdots,n_M}
= &-&2 \({\frac{\pi}{N}}\)^2 \sum_p\sum_{k_p\not=n_p}
\frac{\om^{k_p}\om^{n_p}}{(\om^{k_p}-\om^{n_p})^2}
\({\psi_{n_1,\cdots,k_p,\cdots,n_M} -
\psi_{n_1,\cdots,n_p,\cdots,n_M} }\) \non\\
&-&  \({\frac{\pi}{N}}\)^2
\sum_{pq} \frac{\om^{n_p}\om^{n_q}}{(\om^{n_p}-\om^{n_q})^2}
\psi_{n_1,\cdots,n_M} \non}
The Hamiltonian act on the state of the form
(\ref{EZaa}). Using eq. (\ref{Eex}) given in the
Appendix, one realizes
that the action induced on the polynomials
$\Psi(z)$ is~:
\eqn{
(H_M\Psi)(z) &=& \({\frac{\pi}{N}}\)^2
\({\sum_{p=1}^M z_p\d_{z_p}(z_p\d_{z_p}-N)
 + 4 \sum_{p<q} \frac{z_pz_q}{z_{pq}z_{qp}} }\)\Psi(z) \non\\
&=& \({\frac{\pi}{N}}\)^2
\sum_{p=1}^M \hat D_p (\hat D_p -N) \Psi(z)
\label{ECb} }
In Eq. (\ref{ECb}) one recognizes the Calogero-Sutherland
Hamiltonian at a special value of the coupling constant.
In other words, the spin chain in the $M$-magnon
sector has been mapped on the $M$-body Calogero problem.
The last equality in (\ref{ECb}), gives the energy
of a multiplet $\{m_p\}$:
\eqn{
E(\{m_p\}) = \sum_p\ \({\frac{\pi}{N}}\)^2 m_p (m_p-N)
\non}
This completes the proof of the spectrum.

\section { Wave functions and statistics.}
Only the wave functions of the h.w. vectors
are relevent since those of their descendents are obtained
by recursive action of the transfer matrix.
We now show how recent results on the Calogero models can be
used to find explicit expressions for these wave functions.
The latter are based on the construction of operators,
which we denote by $\La_M$ in the $M$-magnon sector,
intertwining the Calogero Hamiltonian and the free Hamiltonian:
\eqn{ H \ \La_M = \La_M \Delta \quad {\rm with}\quad
\Delta = \sum_{p=1}^M\ z_p\d_{z_p}(z_p\d_{z_p}-N)
\label{EEaa} }
These intertwiners were defined in \cite{Heck,Veselov}.
One of their definitions is the following Vandermond
determinant of the operators $D_p$~:
\eqn{
 \La_M = \sum_{\sig\ {\rm perm.}} {\rm sign}(\sig)\
D^{M-1}_{\sig_{M-1}} \cdots D^2_{\sig_2}D_{\sig_1}
\non }
In this formula, it is understood that $\La_M$ is acting
on antisymmetric functions.
For example, for two magnons: $\La_2=z_1\d_{z_1}-z_2\d_{z_2} -
\frac{z_1+z_2}{z_1-z_2}$.

The operators $\La_M$ are antisymmetric. Therefore,
the symmetric wave functions are obtained by
acting first with the antisymmetrizer on the plane
waves $z_1^{m_1}\cdots z_M^{m_M}$, and then with $\La_M$:
\eqn{
\Psi(z_1,\dots,z_M)= \La_M\ \BL {\rm Det}( z_p^{m_q} )_{pq}\ \BR
\label{EEe} }
It is easy to check that the wave functions (\ref{EEe}) are symmetric
polynomials vanishing at coincident points. If the rapidities
are such that $1\leq m_p\leq (N-1)$, these polynomials have
degree less than $(N-1)$ and satisfy the condition (\ref{EDe}).
They thus are the wave functions of the h.w. vectors.
In other words, since the plane waves $z^m$ are the wave functions
of the one-magnon h.w. vectors, the operator ${\cal D} = \La_M \circ
{\rm Det}$ map tensor products of $M$ one-magnon states into
$M$-magnon states:
\eqn{
 \ket{m_1}\otimes\cdots\otimes\ket{m_M} \
{\buildrel {\cal D} \over \longrightarrow}\ \ket{\{m_1,\cdots,m_M\}}
\label{EEd} }
This map is a generalization of the Slater  determinant in the sense
that it implements the rapidity selection rules: if two
of the rapidities $m_p$ and $m_q$ are either equal or differ by
a unit, then the resulting wave function vanish identically.
The fact that the result vanish if two of the rapidities coincide
is obvious from the definition, while the fact that it vanishes if they
differ by a unit results from an explicit check on the two-magnon
case (which has all generality thanks to the symmetry of the
wave functions and the recursive definition  of $\La_M$ given
in \cite{Veselov}).

\section{The spinon description.}

The magnons are the excitations over the ferromagnetic vacuum;
the excitations over the antiferromagnetic vacuum are conveniently
described in terms of spinons.

For $N$ even, the antiferromagnetic vacuum corresponds to
the alternating sequence of symbols $010101\cdots010$.
Its rapidities sequence is $\{m_j^0=2j-1\}_{j=1}^{N/2}$.
The excitations are obtained by flipping and moving
the symbols $0$ and $1$. We classify the sequence by their number
$M$ of rapidities. The spinon number $N_{sp}$ of a sequence is then defined by
$M=\frac{N-N_{sp}}{2}$. Since $M$ is an integer, $(N-N_{sp})$ is always
even. The spin $S_z$ of the Yangian highest weight vector of the
sequence is $S_z=\half(N-2M)=\frac{N_{sp}}{2}$.

A sequence of rapidities $\{m_j;j=1,\cdots,M=\frac{N-N_{sp}}{2} \}$,
in the $N_{sp}$ sector, can be decomposed into  $(M+1)$ elementary
motifs. As we defined them in Sect.1, an elementary motif is a series
of consecutive $0$. We will think
about the elementary motifs as the possible orbitals for spin
half objects, which are called spinons. At fixed $N_{sp}$, there
are $(1+\frac{N-N_{sp}}{2})$ orbitals. To a spinon in the $j^{th}$
orbital, with $j=0,\cdots,\frac{N-N_{sp}}{2}$, we assign a
momemtum $k=\frac{2\pi j}{N}$. Thus, the spinon momemta vary
from zero to $k_0=\frac{2\pi}{N}\({\frac{N-N_{sp}}{2}}\)$.
By convention, a sequence of rapidities $\{m_j\}$ corresponds
to the filling of the $(1+\frac{N-N_{sp}}{2})$ orbitals with
respective occupation numbers $n_j=n_j^++n_j^-= (m_{j+1}-m_j-2)$,
with $m_0=-1$ and $m_{M+1}=N+1$.
The length of the $j^{th}$ elementary motif is then $Q_j=n_j+1$.
By construction, the total occupation number is the spinon number~:
\eqn{
\sum_{j=0}^{\frac{N-N_{sp}}{2}}\ (n_j^++n_j^-)= N_{sp}
\label{ESa} }
Since an elementary motif of length $Q$ corresponds
to a spin $\frac{Q-1}{2}$ representation of $su(2)$,
the full degeneracy of the sequences is then recovored by
assuming that, at fixed spinon number, the spinon behaves as
bosons. Notice that this property is specific to the $su(2)$
spin chain.

The spinons are not bosons but ``semions" since the number of
available orbitals varies with the total occupation number
\cite{HaF}.  In particular, the spinons are always created by pairs.

The energy of a collection of $N_{sp}=N-2M$ spinons is~:
\eqn{
E-E_{vac}= E_0(M)&+&
\sum_j\sum_\sig 2(M-j)(M-\frac{N}{2}+j) n_j^\sig \label{ESb}\\
&+& \sum_{j,j'}\sum_{\sig\sig'}
 (M-{\rm sup}(j,j')) n_j^\sig n_{j'}^{\sig'} \non}
with $E_0(M)=\inv{3}M(M-1)(4M+1)-(N-1)M^2$.

The low energy, low temperature, behavior is classified
\cite{cardy} as the level one $su(2)$ WZW conformal field
theory. In the spinon description the states consist of
semi-infinite sequences of symbols $0$ and $1$. The
two primary states, which correspond to the two
integrable representations of the $su(2)$ current algebra
at level one, are the vacuum, with sequence
$010101\cdots$ and the spin half primary, with sequence
$0010101\cdots$. The excited states are given by finite
rearrangement of the primary sequence.
The Virasoro generator acts as $L_0= \sum_j(m_j^0-m_j)$, where
$\{m^0_j\}$ are the primary sequences. Its spinon representation is~:
\eqn{
L_0= \({\frac{N_{sp}}{2}}\)^2 + \sum_{j\geq 0}\ j(n_j^++n_j^-)
\label{ESc}}
with $N_{sp}=\sum_j (n_j^++n_j^-)$, the total spinon number.
The excitations over the vacuum possess an even number of spinons,
wheras those over the spin half primary contain an odd
number of spinons.

\section{The thermodynamics.}

Following ref.\cite{Ha91}, the thermodynamics  can be derived
from the spinon description
using methods similar to the thermodynamic Bethe ansatz \cite{YY}.

First, we consider the system with a fixed spinon density
$D_{sp}=\frac{N_{sp}}{N}$. In the $N\to\infty$
limit, the spinon orbitals are labeled by momenta $k$
continuously varying from zero to $k_0=\frac{2\pi}{N}
\({\frac{N-N_{sp}}{2}}\)
=\pi(1-D_{sp})$. Let $n_\pm(k)$ be the spinon occupation
numbers of the $k^{th}$ orbital. By definition, they satisfy~:
\eqn{
\sum_\sig\ \int_0^{k_0}\frac{dk}{2\pi}\ n_\sig(k) = D_{sp}
\label{ETa} }
In the continuum limit,
the energy per unit of length is~:
\eqn{
\frac{E}{N} \equiv \CE(D_{sp};n_\pm(k)) =
&& \CE_0(D_{sp}) + \sum_\sig \int_0^{k_0}\frac{dk}{2\pi}\
\ep_0(k) n_\sig(k) \non\\
&&+ \sum_{\sig\sig'}
\int_0^{k_0}\frac{dkdk'}{(2\pi)^2} V(k,k') n_\sig(k) n_{\sig'}(k')
\label{ETb} }
with $\CE_0(D_{sp})= (\frac{\pi}{2})^2(1-D_{sp})^2(\frac{1+2D_{sp}}{3})$,
$\ep_0(k)=\half(k_0-k)(k_0+k-\pi)$
and $V(k,k')=\frac{\pi}{2}(k_0-{\rm sup}(k,k'))$. In each orbital the spinons
behave as bosons, therefore the entropy of a configuration of
$n_\pm(k)$ spinons is~:
\eqn{
\frac{S}{N} \equiv \CS(D_{sp};n_\pm(k)) =
\sum_\sig \int_0^{k_0}\frac{dk}{2\pi}\
\({ (n_\sig(k)+1)\log(n_\sig(k)+1) - n_\sig(k)\log n_\sig(k)}\)
\label{ETc} }
The free energy per unit of length is~:
\eqn{
\frac{F}{N} = \CF = \CE - T\CS - h\CM
\label{ETd} }
with $h$ the exterior magnetic field and $\CM=\int_0^{k_0}\frac{dk}{2\pi}
(n_+(k)-n_-(k))$ the magnetization.

At fixed spinon density, the thermodynamic equilibrium is determined
by minimizing (\ref{ETd}) with respect to the variation
of the spinon occupation numbers subject to the constraint
(\ref{ETa}). This gives~:
\eqn{
n_\pm(k) = \inv{e^{\beta(\ep(k)\mp h - A)} - 1}
\label{ETe}}
where $A$ is the Lagrange multiplier and $\ep(k)$ is
the dressed energy defined by~:
\eqn{
\ep(k)= 2\pi\frac{\de \CE}{\de n_\sig(k)}
= \ep_0(k) + 2 \sum_{\sig'}
\int_0^{k_0}\frac{dk'}{2\pi} V(k,k') n_{\sig'}(k')
\label{ETf} }
At fixed density, the Lagrange multiplier is determined
from the constraint (\ref{ETa}). This completely
specifies the thermodynamics of the spinon gas.

The spin chain corresponds to a spinon gas of arbitrary
density; i.e. the spinon chimical potential
$\mu=\frac{\d \CF}{\d D_{sp}}$ is zero.
Minimazing the free energy with respect to the density
fixes $A$ to be $\beta A= -\log(2\cosh(\beta h))$.
The constraint (\ref{ETa}) then gives the mean
density $\bar D_{sp}$.

The coupled eqs. (\ref{ETe},\ref{ETf}) can be solved. Deriving twice
eq.(\ref{ETf}) with respect to $k$ gives~:
\eqn{
\frac{\d^2 \ep(k)}{\d k^2} = -\({1+\half \sum_\sig n_\sig(k) }\)
\label{ETg} }
with the boundary conditions: $\ep(k_0)=\ep(0)=0$, and
$\ep'(0)=-\ep'(k_0)=\frac{\pi}{2}$.
Eq.(\ref{ETg}) is integrated by introducing the dressed
momenta $p=\frac{d \ep}{d k}$. It varies from $-\frac{\pi}{2}$
to $\frac{\pi}{2}$, and it satisfies $\frac{d p}{d k} =
\frac{d^2 \ep}{d k^2}=\frac{d (p^2/2)}{d \ep}$.
As function of $p$, the occupation number are then given by~:
\eqn{
n_\pm = \exp\[{\beta(\eta \pm \ep_{dr}\pm h)}\]
\label{ETh} }
with
\eqn{
\eta(p)=\half\[{p^2- (\frac{\pi}{2})^2}\]
\quad{\rm and}\quad
\exp(\beta\eta) = \frac{\sinh(\beta\ep_{dr})}{\sinh(\beta h)}
\label{ETj}}
In the limit $h\to 0$, the dressed energy is $\ep_{dr}\sim he^{\beta\eta}$.

The free energy is found by integrating the thermodynamical relations~:
\eqn{
\({ \frac{\d \CF}{\d T}}\)_h = - \CS
\quad {\rm and}\quad
\({\frac{\d \CF}{\d h}}\)_T = - \CM }
The result is the following simple answer~:
\eqn{
\CF = -\ T\int_{-\frac{\pi}{2}}^{\frac{\pi}{2}}\frac{dp}{\pi}\
\log\[{\frac{\sinh(\beta(\ep_{dr}+h))}{\sinh(\beta h)} }\]
\label{ETl}}
Notice that this is the free energy of a gas of non-interacting
particles which, in the limit $h\to 0$, have energies given
by $\eta(p)$.

\section{Appendix.}
In this Appendix we prove the eqs.(\ref{EDh},\ref{EDhh}).
First we compute the action of  $T_{--}(x)$ on these magnon states.
We recall that $T_{--}(x)$
can be written as $T_{--}(x)=1+\sum_{i,j} \({\inv{x-L}}\)_{ij}P^-_j$,
where $P^-_j$ is the projector on the spin $(\sig_j=-)$ acting on
the $j^{th}$ spin only.
The projector $P^-_j$ on the $M$-magnon states (\ref{ECa}) gives states
with one spin down marked of the form~:
\eqn{ P^-_j\ket{\Psi}\equiv
\ket{\Psi_j}= M \sum_{n_2,\cdots,n_M}\psi_{j;n_2,\cdots,n_M}
\sig_j^-\sig^-_{n_2}\cdots\sig^-_{n_M}\ket{\Om}
\label{EZdd} }
They corresponds to polynomials $\Psi(z_1|z_2,\cdots,z_M)$ symmetric
in $z_2,\cdots,z_M$ with the point $z_1$ distinguished.
To evaluate the action of $\sum_{i,j} (L^n)_{ij} P^-_j$,
we remark that on this class of states, $L_{ij}$ acts as follows~:
\eqn{
\sum_k\ L_{jk}\ket{\Psi_k}= \ket{(L\Psi)_j}
\label{EDf} }
with,
\eqn{
(L\Psi)_{j;n_2,\cdots,n_M}=
\sum_{k} \ \hat\theta_{jk} \psi_{k;n_2,\cdots,n_M}
+\sum_{q=1}^M \ \hat\theta_{jn_q} (K_{1q}\psi)_{j;n_2,\cdots,n_M}
\label{EDg} }
where $\hat\theta_{jk} = \om^j/(\om^j-\om^k)$, and $K_{1q}$
permutes the indices in position $1$ and $q$.
Therefore, $\ket{(L\Psi)_i}=\sum_j(L^n)_{ij}P^-_j\ket{\Psi}$
can be recursively computed. Then,
\eqn{
\sum_{i,j}(L^n)_{ij}P_j\ket{\Psi} = M \sum_i\ket{(L^n\Psi)_i}
\label{EZgg} }
is obtained by symmetrizing the wave
function coefficients of $\ket{(L^n\Psi)_i}$ in all its indices.
I.e. its wave function coefficients, denoted $(L^nP\Psi)_{n_1,\cdots,n_M}$
are ~:
\eqn{
(L^nP\Psi)_{n_1,\cdots,n_M} = \({ 1 + \sum_{p\not= 1} K_{1p} }\)
(L^n\Psi)_{n_1;n_2,\cdots,n_M}
\label{EZhh} }

We now translate this action on the wave function coefficients
into an action on the polynomials.
We recall that in the basis of polynomials $Q_k$ in one
variable of degree $(N-1)$ specified by $Q_k(\om^n)=\de^n_k$,
the matrix elements of the derivatives are:
\eqn{
( z\d_z - \frac{N+1}{2})Q_j(z) &=&
- \sum_{k\not= j}\ \frac{\om^j}{\om^j-\om^k} \ Q_k(z)
\label{Eex}\\
z\d_z(z\d_z-N)Q_j(z) &=&
-2\sum_{k\not= j} \frac{\om^j\om^k}{(\om^j-\om^k)^2}
(Q_k(z)-Q_j(z)) \non}

The differentials $D_p$, introduced in eq.(\ref{ECd}),
 acts on polynomials of the form $(\ref{EZaa})$;
i.e. they preserve the form of these polynomials. Moreover,
by comparing the formula, we recognize in eq.(\ref{EDg}) the operator
$\({D_1-\frac{N+1}{2}}\)$. Since the operators $D_p$
are covariant by permutation of the coordinates,
the polynomial of $(L^nP\Psi)$ is given by acting with
$\sum_p \({ D_p-\frac{N+1}{2}}\)^n$ on $\Psi$.
Resuming all the contributions, we obtain~:
\eqn{
T_{--}(x)\  \Psi(z_\cdot)\
= \({1 + \sum_{p=1}^M\inv{x+\frac{N+1}{2}-D_p} }\) \Psi(z_.)
\non}
The action of $T_{-+}(x)$ on these subclass of magnon states
can be computed using the same methods. Its action is
given by eq.(\ref{EDhh}).
It is remarkable that the differentials $D_p$ appear naturally in
the study of the spin transfer matrix.



\end{document}